\RequirePackage{fix-cm}
\documentclass[twocolumn]{svjour3}          
\smartqed  
\usepackage{lineno,hyperref,amssymb,url,graphicx,array,amsmath,textcase,bm,float}
\usepackage{graphicx}
\usepackage{cite}

\begin{document}

\title{Nanosecond laser ablation of Si(111) under an aqueous medium}


\author{Erkan Demirci         \and
        Elif Turkan Aksit Kaya \and
        Ramazan Sahin
}


\institute{Erkan Demirci and Elif Turkan Aksit Kaya \at 
              Informatics and Information Security Research Center, Tubitak-Bilgem 41470, Turkey \\
           \and
           Ramazan Sahin \at
              Department of Physics, Faculty of Science, Akdeniz University 07058, Turkey \\
               \email{rsahin@itu.edu.tr}                         
}

\date{Received: date / Accepted: date}

\maketitle

\begin{abstract}
Nanosecond laser ablation of p-type Si(111) wafer is presented under an ambient, a water and a glycerin conditions. Effects of pulse energy, number of pulses and type of environment are systematically analyzed with an Optical Profilometer, a Scanning Electron Microscope and an Atomic Force Microscope. Obtained results are compared with $Zemax^{\circ}-EE$ simulations. Self-assembly nano-ripple formation has been observed in an ambient condition. However, these ripples and heat affected zone (HAZ) area disappeared in aqueous medium which enhances the quality of ablation. Moreover, no remarkable oxidation was obtained in ablation zone even in an ambient condition based on the SEM-EDX analysis. Ablation profiles and their depths for different environments are compared with false color technique. Although ablation diameters are comparable each other, their depths are small enough in aqueous medium to control depth profile of damaged area. According to our results, best ablation profiles (precise edges without HAZ) are achieved under an aqueous medium especially in glycerin conditions both in low and high pulse energy regime. 
\end{abstract}

\section{Introduction} \label{sec:intro} 
Since the method does not require any chemical process or special environments such as vacuum, surface modification without touching the target at micro- and nano-scale dimensions has attracted much attention since the higher intensity pulsed laser systems were established \cite{liu_1997,karabutuv_2006,pronko_1995,simon_1996}. The laser material-interaction and its resulting effect depends naturally on the pulse duration. In ultrafast regime (the laser-material interaction time is in the order of picosecond or femtosecond), a plasma of hot-electrons occurs and energy of this plasma can be enough for breaking of chemical bonds yielding an ablation \cite{mazur_2006, trtica_2007}. On the other hand, when the pulse duration is comparable with heat diffusion time, the laser energy is transferred to target material causing local heating and raising temperature before vaporization \cite{zhou_physical_2011}. Therefore, the mechanism in the thermal regime depends on both material properties such as thermal conductivity, optical absorption and laser parameters such as wavelength, pulse duration, laser fluence  \cite{jellison_1982,korner_1996,ulrich_2008}. 
As a semiconductor material, Silicon (Si) has always been used in microelectronics, sensors, solar cell technology and biotechnological applications \cite{yates_1998}. In order to open insight to new technologies or miniaturize and increase the performance of Si based devices, patterning their surfaces at micrometer scales are quite important \cite{martin_2014, lee_analysis_2010, zuev_fabrication_2011, kim_2013}. Moreover, better precision of such structures is an important requirement in laser-based fabrication processes \cite{chong_2010,malinaskus_2016}. 

Nanosecond lasers are relatively cheaper than ultrafast lasers and yield smaller Heat Affected Zone (HAZ) area than continuous or long-pulse duration (e.g. $\mu$s) lasers. Due to the optical properties of Si, most of the studies are focused on nanosecond laser ablation in visible or UV regime to increase absorption of incoming energy \cite{watanabe_2010, karnakis_2006, yoo_2000}. However, a fundamental wavelength of laser (e.g. 1064 nm) does not require a harmonic generation which yields a better alternative for straightforward and cheaper fabrication.

Time resolved analysis in nanosecond regime reveals that the Si sample is first heated and melted before vaporization due to laser irradiation, respectively \cite{zhou_physical_2011}. These cause in inevitable debris formation and re-solidification of melted materials due to plasma expansion or HAZ area surrounding the focal diameter in nanosecond regime where an additional surface treatment is required to remove debris or clean the damaged area. Nowadays, many researchers focused on laser ablation of materials in aqueous medium \cite{watanabe_2010,jimenez_2009,leksina_2016,karimzadeh_2009} to remove residual effects. 

In this study, the ablation mechanism of nanosecond laser ($\lambda_{laser}$=1064 nm) on Si(111) in an aqueous medium is investigated through systematical single pulse ablation of in the ambient, the water and the glycerin conditions and characterization of damaged areas.
\section{Experiments and Calculation} \label{sec:ablation} 
The p-type (doped with boron) Si(111) wafers with a thickness of 510 $\mu$m were used in this study. First, these wafers of a diameter of 76.2 mm were cut into small pieces before washing them with ultrasonic cleaner in an acetone, an alcohol and a pure-water, respectively. The sheet resistance of the small pieces was equally measured as 15 $\Omega$/sq with a four-point-probe. 

The electro-optic Q-switch Nd:YAG laser system (Ekspla NL230) generates 5.5 ns pulses at the fundamental mode ($\lambda_{laser}$=1064 nm). The laser system can be controlled to operate in a single- or a multi-pulse mode (up to 100 Hz). The pulse energy can be adjusted between 5 and 80 mJ by Q-switch delay adjustment. The pulse to pulse energy instability was measured as 0.27$\%$ for the maximum power. The polarization of the laser beam is kept constant as horizontal throughout the experiments. The beam divergence is less than 0.8 mrad and the $M^2$ value is below 2.5. The output diameter of the laser beam is 4.5 mm and nearly Gaussian in the far field. 

\begin{figure}[H]
	\centering
		\includegraphics[width=0.45\textwidth]{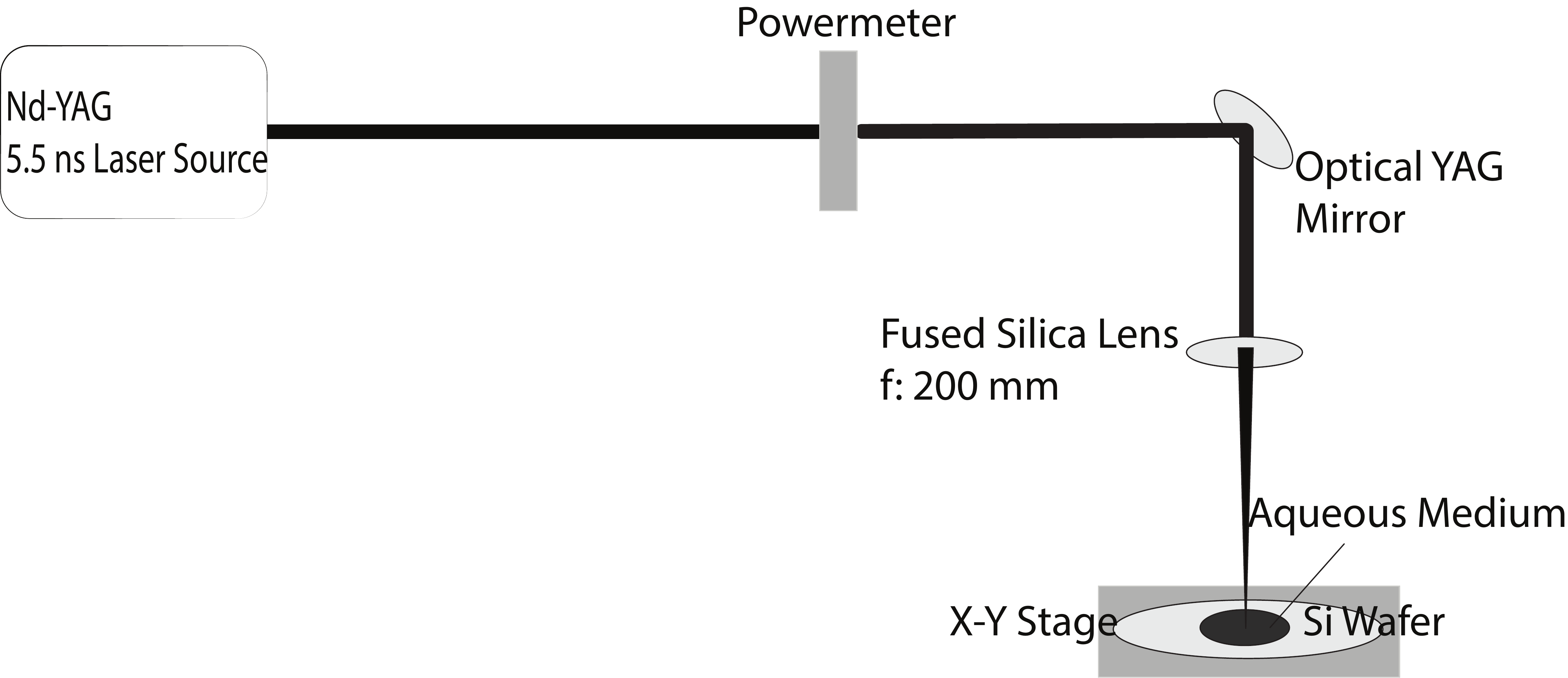}
			\caption{Experimental setup for ns laser ablation of Si wafers in different medium}
	\label{fig_setup}
\end{figure}

Experimental setup is shown in Fig. \ref{fig_setup}. Ablation experiments were conducted in the far field. The optical YAG mirror (Edmund optics 33-076) is used to send the laser beam perpendicularly before focusing the beam on the sample surface with a plano-convex lens (Thorlabs, LA4894) made of fused silica in which the focal length of lens is 200 mm. The experimental setup is adjusted so that the distance between lens and the sample surface (z-direction) is kept constant.

The laser system is located in ISO 8 class (100000 class) clean-room. The temperature is stabilized at 18 $^{\circ}C$ and humidity is 50 $\%$. The samples were positioned at the focal point of lens on translation stage which can move in x- and y-directions. The experiments were conducted under three different environments for a comparison; the ambient, the water and the glycerin. The 2 mm thick water or glycerin (V=2400 $mm^3$) layer was placed on the sample.

For the characterization of ablated samples, Bruker ContourGT-K 3D Optical Microscope (OM) was used. The lateral resolution of the microscope is 0.5 $\mu$m while vertical resolution is less than a nanometer. Diameters of ablation zones and Heat Affected Zones (HAZ) can be measured from obtained OM images. In addition, Scanning Electron Microscope (SEM), JEOL-JSM 6335F was used in Secondary Imaging Mode (SEI). Besides, an Atomic Force Microscope (AFM) from Nanomagnetics Instruments is used to obtain surface topography.

Laser beam radius is theoretically calculated first by using equation in \cite{self_focusing} regardless of surrounding medium. For our experimental setup and parameters given above, the beam diameter at focus is calculated as 156 $\mu m$. Then, $Zemax^{\circ}-EE$ Optical Design program was used to estimate beam radius at focus in different medium. In addition, the beam diameter at focus was obtained experimentally by plotting the square of ablation diameter ($D^2$) versus the laser fluence explained as follows. 

Laser beam radius is calculated by using $Zemax^{\circ}-EE$ Optical Design Program in the sequential mode. $n_{air} = 1$, $n_{water} = 1.3262$ and $n_{glycerin} = 1.4631$ are the input parameters for the refractive index of each medium. Moreover, the thickness of the liquid layers is assumed as 2 mm. First, experimental setup is designed in the software with laser (wavelength, $M^2$ value, pulse duration, beam radius at laser output) and lens ($n_{lens}$, focal distance) parameters. Abbe numbers are 55.8 and 54.6 for water and glycerin conditions, respectively. Since we just used fundamental mode of laser cavity ($\lambda_{laser}$=1064 nm), the deviation of the partial dispersion is taken 0.0 for all materials. All of these input parameters yield focal diameter of 103.6 $\mu m$ in air, 103.07 $\mu m$ in water and 103.06 $\mu m$ in glycerin after calculation of paraxial Gaussian beam analysis in the $Zemax^{\circ}-EE$. The beam radius value in water or glycerin is a little bit smaller than that in an ambient condition due to the self-focusing effect. However, since the liquid layers are so thick (2 mm) compared to the focal distance of lens (200 mm), beam diameter at focus would not be changed much. 

For a Gaussian laser beam, fluence distribution ($F(r)$) in the transverse plane is given by Eq. \ref{gauss_fluence};

\begin{equation}
F(r) = F_0 exp(-2\frac{r^2}{w^2_{0}})
\label{gauss_fluence}
\end{equation}

where r is the radial coordinate, $w_0$ is the beam radius at beam waist and $F_0$ is the maximum laser fluence at the center. The maximum laser fluence is equal to $2E_{0}/\pi w^2_{0}$ where $E_0$ is the pulse energy. From these equations, one can evaluate the ablation diameter (D). 

\begin{equation}
D^2 = 2w^{2}_{0} \left[ln(F_{0})-ln(F_{th})\right]
\label{ablation_diameter}
\end{equation}

By plotting the logarithm of applied laser fluence and square of ablation diameter, ablation threshold ($F_{th}$) and beam radius can be evaluated from intersection and slope of the graphic, respectively. However, this is generally used to obtain ablation threshold of any kind of materials in ultra-fast regime (femtosecond or picosecond pulsed lasers) \cite{mannion_2004, sivakumar_2003}. Ablation mechanism in ultra-fast regime is explained totally by a non-thermal regime. On the other hand, pulse duration is much longer in nanosecond regime to allow enough amount of energy to be transferred into the target material for local heating. This yields damaged area surrounding of laser spot on sample which is larger than focal diameter of laser beam. Therefore, one would not expect logarithmic behavior of ablation diameter with laser pulse fluence in air conditions. Since the ablation regime is quite different, especially, from the ultrafast laser-target material interaction in nanosecond pulsed laser regime, an approach to determine ablation threshold of target material was developed by \cite{karimzadeh_2009} for liquid environment that we used equation in \cite{karimzadeh_2009} to calculate experimental focal beam diameters for liquid environments. 

\section{Results and Discussion}
\label{sec:evaluation}
The ablation experiments were organized in three parts; (i) under the ambient condition, (ii) in the water and (iii) in the glycerin conditions. 

\begin{table}[h]\footnotesize
\begin{tabular}{ | c | c | c |}
\hline 
Properties                   &Water       &Glycerin     \\ \hline 
Refractive Index (n)         &1.433       &1.475        \\ \hline 
Thermal Conductivity (W/m.K) &0.609       &0.285        \\ \hline 
Viscosity (Pa.s)             &$10^{-3}$   &1.412        \\ \hline 
Density ($g/cm^{3}$)         &$10^{-3}$   &1.261        \\ \hline 
Color                        &colorless   &colorless    \\ \hline 
\end{tabular}
 \caption{Physical properties of water and glycerin.}
\label{table}
\end{table}

Some physical properties of water and glycerin are given in Tab. \ref{table}. Although the refractive index of glycerin is bigger than that of water, thermal conductivity of glycerin is much smaller. For comparison of ablation diameters in the aqueous medium with the ambient condition, the same laser pulse energies were used while observing damaged area with OM and SEM.

\begin{figure}[H]
	\centering
		\includegraphics[width=0.4\textwidth]{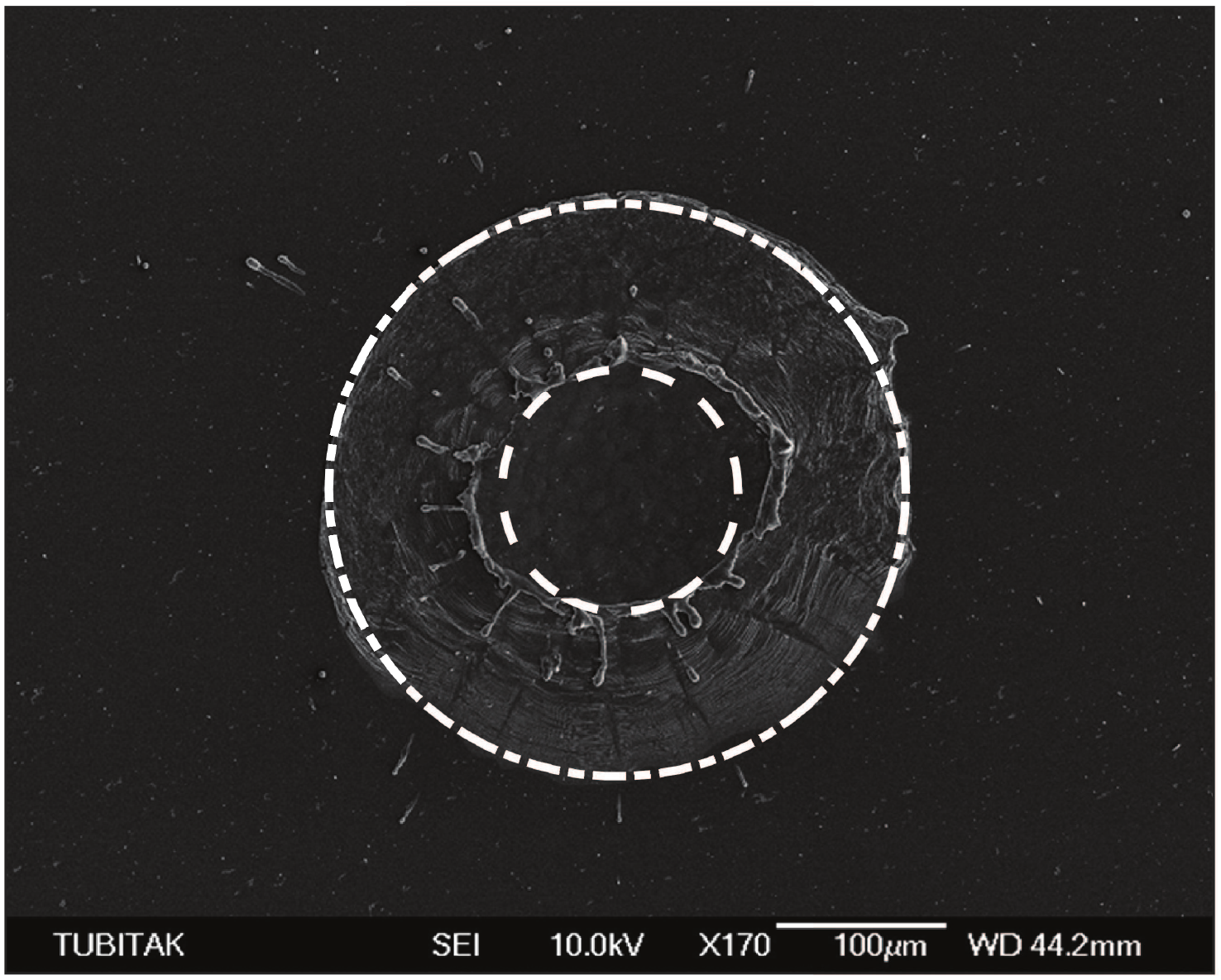}
			\caption{SEM image of an damaged area under ambient condition in which inner dashed ring defines the ablated zone whereas outer ring surrounds the HAZ area.}
	\label{SEM_HAZ}
\end{figure}

Although the sample is almost transparent at 1064 nm, Fig. \ref{SEM_HAZ} shows surface morphologies of a damaged area on the Si(111) surface irradiated by a nanosecond laser at a single pulse ($E_{pulse}$=18 mJ) under ambient conditions. The depth of ablated crater is 13.8 $\mu$m. Neither there is any residual effect nor any obvious structure formation was observed inside ablated zone. On the other hand, most of the ejected molten material from relatively smooth central part has been re-solidified in the peripheral area (HAZ zone). The diameter of the ablated area was in the order of beam focal diameter while the diameter of the damaged area is much bigger than focal diameter due to thermal expansion. Besides, there were some features (recoiling and ripple formation zones) in the HAZ area. We also performed EDX analysis in the laser ablated area and found out that 96.1 $\%$ of the area was composed of Si while only 3.9 $\%$ was Oxygen (which disappears for aqueous medium based on our results) even though ablation experiments were conducted on the ambient condition.

\begin{figure}[H]
	\centering
		\includegraphics[width=0.45\textwidth]{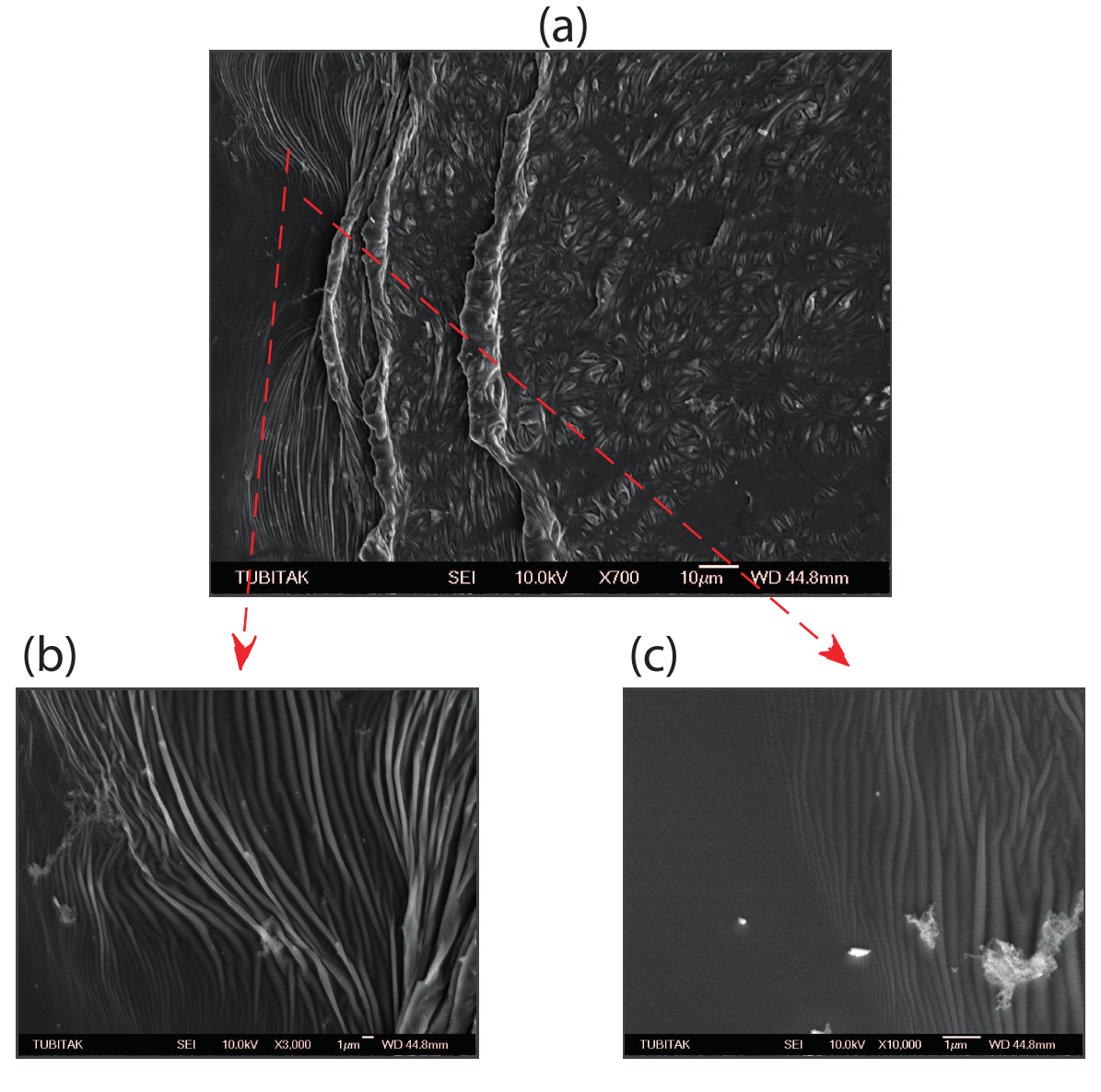}
			\caption{(a) SEM image of an edge between ablated and HAZ area (b) and (c) are the selected areas from (a). The pulse energy is 50 mJ for fabrication of this crater formation in an ambient condition.}
	\label{riple_SEM}
\end{figure}

In order to discover details of the surrounding area of the ablated zone, a higher magnification SEM images were obtained in damaged area. Fig. \ref{riple_SEM} indicates SEM image of edge of ablated zone. Sub-wavelength ripple formation around ablated zone is obvious in the SEM images. Based on our measurement, the ripples are found to be approximately 50 $\mu$m long and their periodicity was around 300 nm. 

\begin{figure}[H]
	\centering
			\includegraphics[width=0.3\textwidth]{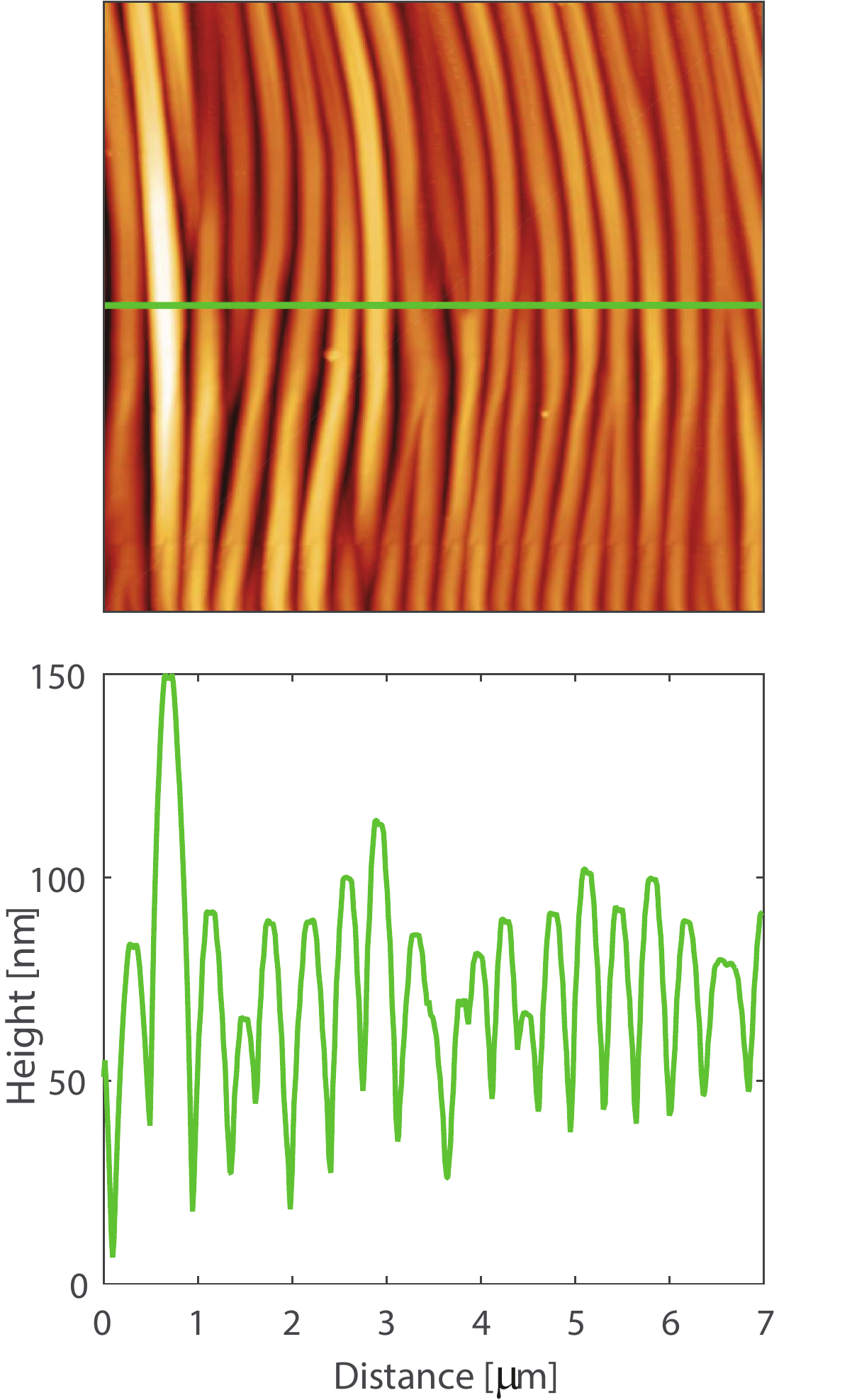}
			\caption{AFM topography image of an self-assembly nanoripples acquired between HAZ and ablated area.}
	\label{AFM_ripple}
\end{figure}

We also conducted on AFM measurements on the ripple formation area. CoreAFM from Nanosurf Ins. in tapping mode was used for obtaining surface topography of Si(111) sample (Fig. \ref{AFM_ripple}). Based on our results, the average height of ripples was around 100 nm. Since we observe these ripples relatively in higher pulse energies we think that the origin of these self-assembly formations is due to stress between borders of HAZ and ablated area due to confinement of melting material to that area. Since the HAZ area is absent in the liquid environments, these ripples could not be fabricated in our ablation experiments for liquid environments.

\begin{figure}[H]
	\centering
			\includegraphics[width=0.45\textwidth]{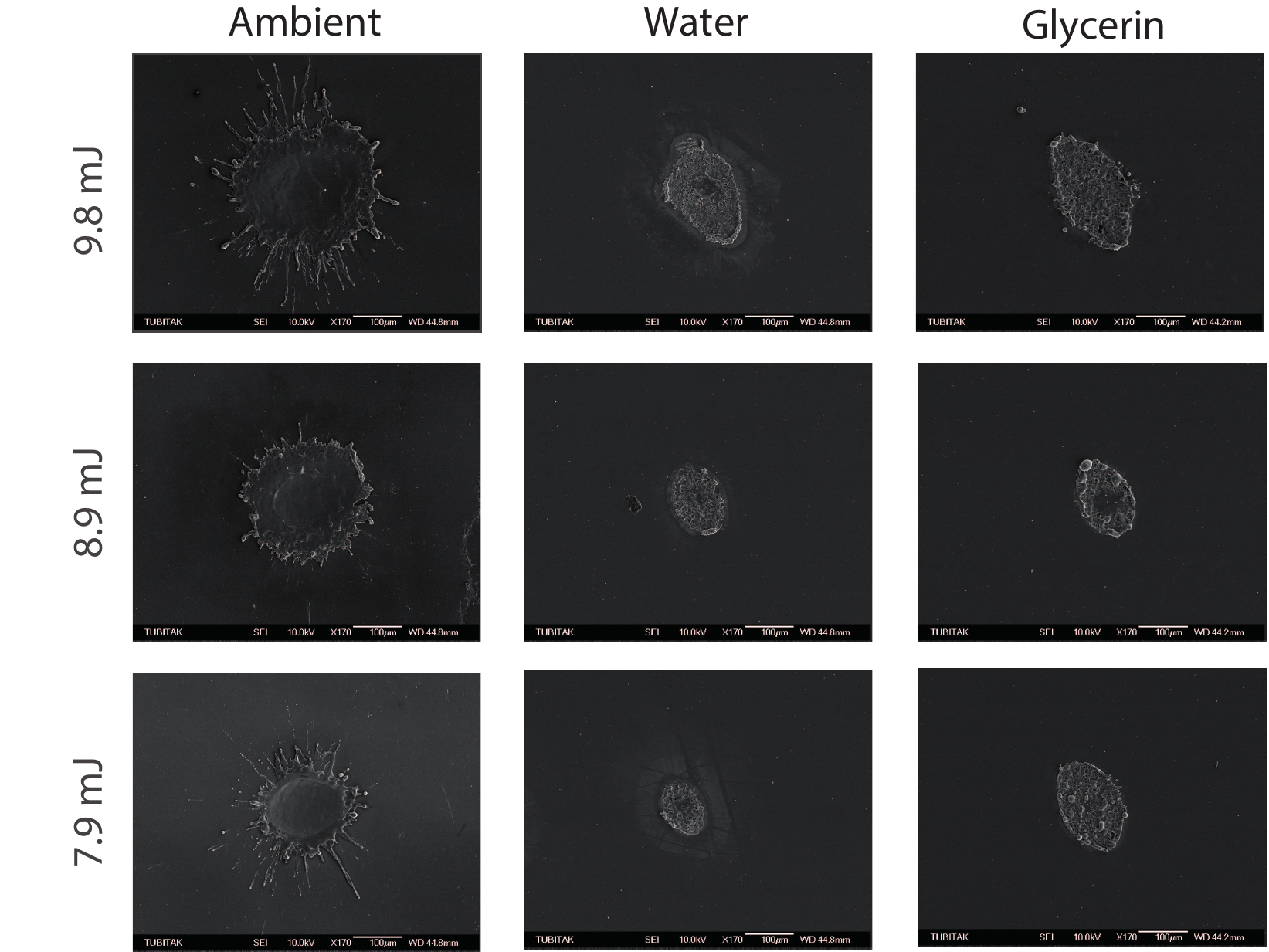}
			\caption{Effect of surrounding medium on the ablation and HAZ areas. The laser pulse energy is in the range of 7.9 and 9.8 mJ}
	\label{air_water_glycerin}
\end{figure}

It was reported in the literature that liquid medium provides more effective cooling and higher optical breakdown threshold than that of air \cite{karimzadeh_2009}. We also performed ablation experiments with equal pulse energies for air, water and glycerin medium.

Fig. \ref{air_water_glycerin} shows SEM images of damaged areas for variety of pulse energies and different environments. As the pulse energy decreases the diameter of damaged area also gets smaller literally. Moreover, both the diameter of a HAZ area and residual effects (redeposition of melting material and debris formation) could be reduced much in ablation experiments for water medium. The thermal conductivity of glycerin is smaller than that of water and hence, it is less prone to HAZ area formation resulting in ablated area being smaller than laser focal diameter. In addition, the dense medium such as water and glycerin limits the heat convection (heat transfer by mass motion). Therefore, we achieved in production of precise ablated spots almost completely and neither HAZ area nor any residual effects (re-solidification or debris formation) could be observed in our experiments in the glycerin.

\begin{figure}[H]
	\centering
	\includegraphics[width=0.5\textwidth]{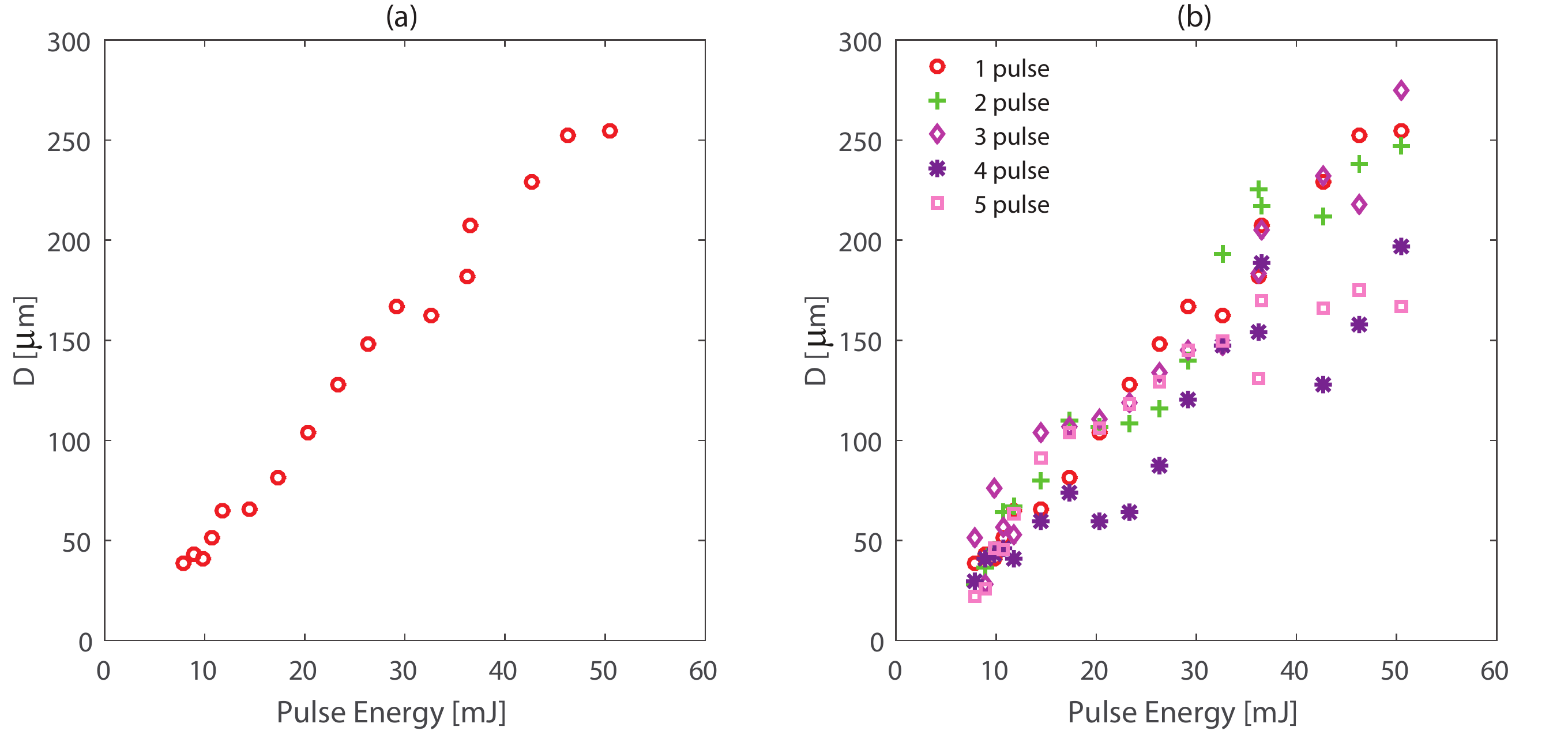}
			\caption{(a) The effect of laser pulse energy on the ablation diameter under ambient conditions and (b) the effect of number of pulses at lower repetitions on the ablation diameter}
	\label{energy_diameter_air}
\end{figure}

In order to quantify ablation diameter and obtain pulse energy dependence on the damage formation under ambient conditions systematical experiments were carried out. The laser beam is nearly Gaussian. Therefore, the square of ablation diameter is proportional to natural logarithm of laser pulse fluence in ultrafast laser ablation of materials. Since the Si(111) sample was ablated by nanosecond pulsed laser, the thermal effects (hot plasma model) should be taken into account and the behavior of ablation diameter should be analyzed differently. Ablation diameters measured from both single pulse and multi-pulse experiments are drawn in Fig. \ref{energy_diameter_air} with respect to laser pulse energy. We found that the diameter of ablated zone is linearly dependent on the pulse energy, Fig. \ref{energy_diameter_air} (a). The measurements of ablation diameter in multi-pulse ablation experiments at lower repetition rates follows linear dependence Fig. \ref{energy_diameter_air} (b). Besides, 40 $\mu$m crater diameters on Si(111) surface could be achieved in both single and multi-pulse ablation experiments.

\begin{figure}[H]
	\centering
		\includegraphics[width=0.45\textwidth]{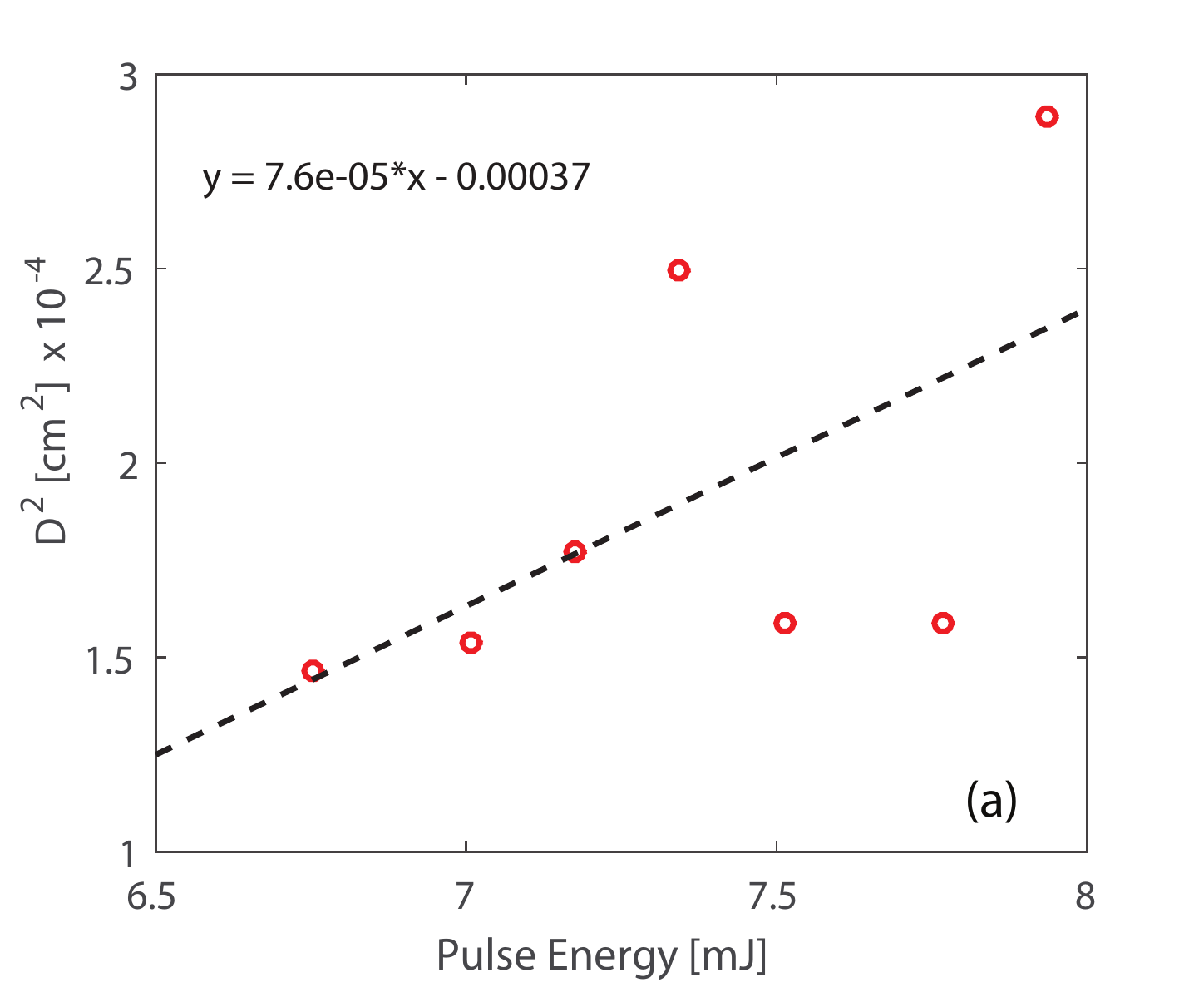}
				\includegraphics[width=0.45\textwidth]{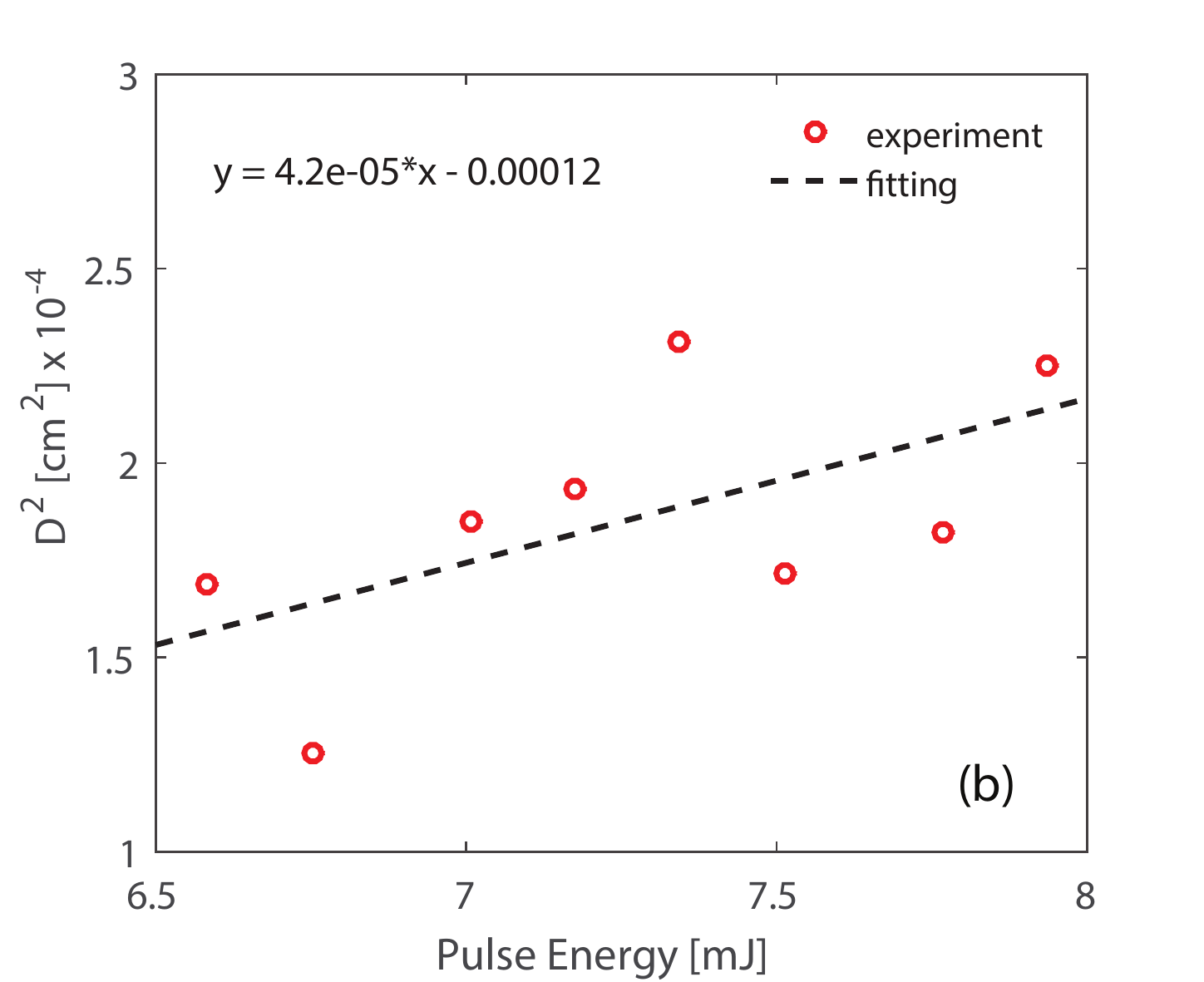}
			\caption{Effect of laser pulse fluence on the square of ablation diameter of Si(111) sample for (a) water and (b) Glycerin environment.}
	\label{energy_diameter_aqua}
\end{figure}

For comparison, we repeated similar experiments with the sam pulse energies in the liquid environments. Results are shown in Fig. \ref{energy_diameter_aqua}. The relationship between square of ablation diameter and natural logarithm of laser pulse energy is used to estimate ablation threshold and focal diameter for liquid environment \cite{karimzadeh_2009}. We used this approach to obtain ablation threshold values for water and glycerin environment by fitting the square of ablation diameter versus the natural logarithm of laser pulse energies. The beam diameter at focal point in aqueous medium could be estimated from the slope of this fitting. 83.6 and 58.9 $\mu$m focal beam radius values were found for water and glycerin environments, respectively. 47.7 and 52.2 J/$cm^2$ ablation threshold values were obtained for water and glycerin environments at single pulse conditions. Although, there is a deviation between theoretically calculated and experimentally estimated focal beam diameter values, they are in the same order. Moreover, the Si sample ablation threshold values are very close to previously reported works \cite{trtica_2007,zhou_physical_2011,tao_time_resolved}. Since we use laser with a wavelength of 1064 nm being much larger than UV and visible wavelengths and carries smaller photon energy, one should expect a little bit higher ablation threshold values in addition to low absorption coefficient of Si sample at this wavelength \cite{karnakis_2006,karimzadeh_2009}.

The control of ablation depth would be sometimes major subject when it comes to surface modification at micrometer dimensions. Before ablation experiments the average roughness of the samples was measured as 0.2 nm with an AFM. In these measurements, the total image sizes taken in tapping mode AFM where the cantilever was excited at its resonance frequency were 100 $\mu$$m^2$.

\begin{figure}[H]
	\centering
		\includegraphics[width=0.45\textwidth]{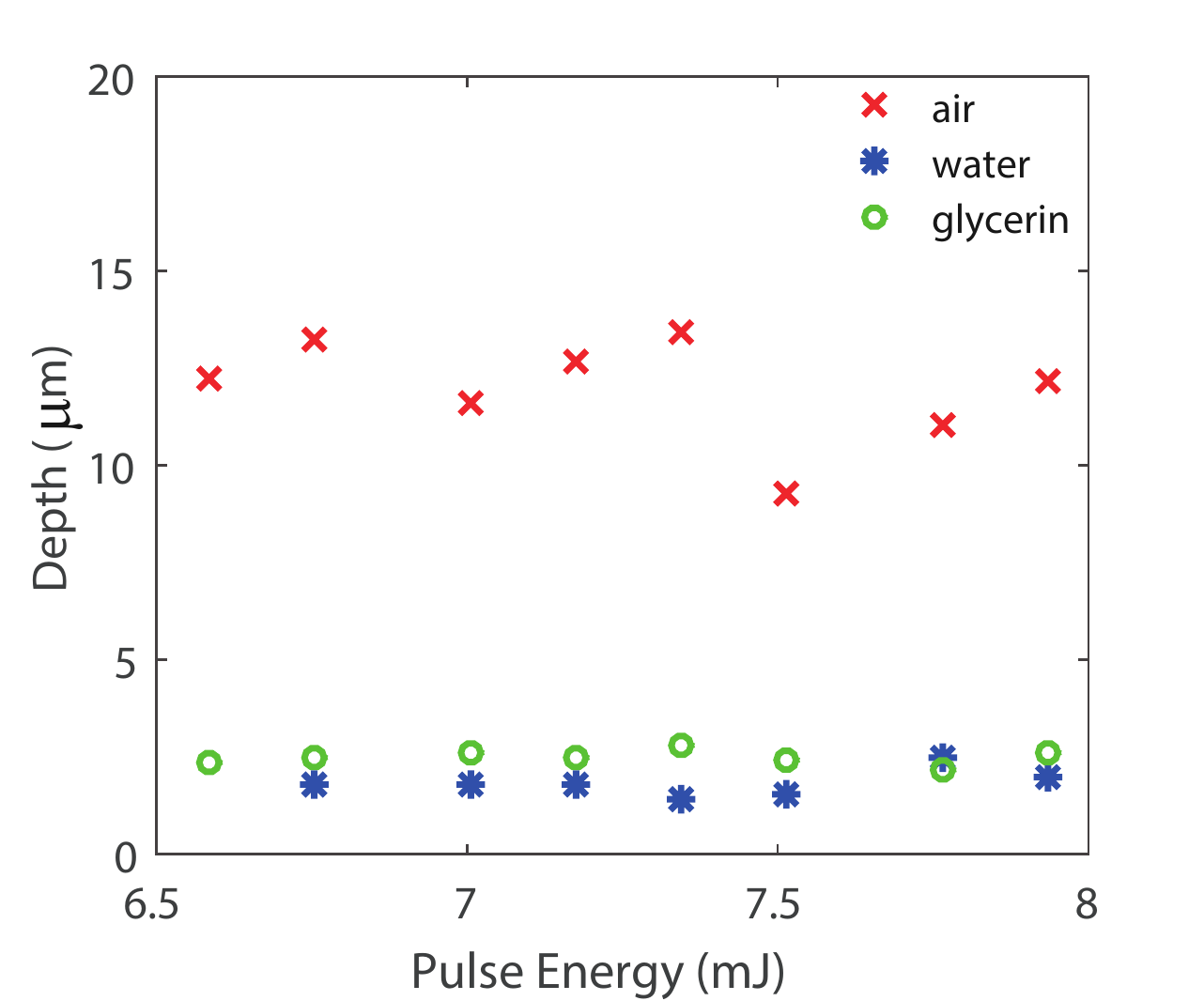}
			\caption{Effect of laser pulse energy on the ablation depth. The crater depth is measured from center of ablated area.}
	\label{energy_depth}
\end{figure}

Single pulse ablation experiments were conducted on Si(111) sample for an ambient, a water and a glycerin surrounding medium. We observed in our experiments and analysis that regardless of type of liquid environment the ablation depth is $\sim$2.5 $\mu$m. However, this depth value reaches $\sim$13 $\mu$m under the ambient conditions, Fig. \ref{energy_depth}. Comparable depths were obtained by the other researchers \cite{trtica_2007, martin_2014} for similar laser wavelengths and pulse durations. We think that the energy is confined to interface of surface of Si(111) sample and liquid for aqueous environment and this liquid behaves as a cooling medium also. Results show that the more control on ablation depth is provided through nanosecond laser-material ablation in an aqueous medium.

\section{Conclusion}
Nanosecond pulsed laser ablation of Si(111) sample was analyzed in an aqueous medium. For the characterization of fabricated structures OM, SEM and AFM were used. Experimental results were compared with theoretical calculations and $Zemax^{\circ}-EE$ simulations. Regardless of surrounding medium $\sim$40 $\mu$m structures could be achieved in Si(111) surface. Our analysis show that HAZ area was much reduced in water environment. On the other hand, HAZ area almost completely disappeared for glycerin condition. Moreover, neither re-deposition of melting material nor debris formation has been observed in the peripheral area of ablated crater which provides better precision for ns laser ablation of Si(111) sample. We found out ablation threshold values of Si(111) sample for liquid environments. Our results are very convenient with the previously reported results. Since we use longer wavelengths and the optical absorption of Si sample is very low at this wavelength, ablation threshold values in our experiments are a little bit higher. Based on our analysis, aqueous medium yields reduced the ablation depth. $\sim$2.5 $\mu$m and $\sim$13 $\mu$m average values were obtained in our experiments. These results and observations suggest that reduced HAZ area could be achieved in the aqueous conditions provided that lower thermal conductivity liquids are employed in ns laser ablation experiments without requiring any liquid flowing.

\begin{acknowledgements}
R.Sahin thanks The Scientific Research Project Unit of Akdeniz University (Project Numbers: FAY-2017-2530, FYD-2017-3058, FYD-2017-3057) for partial support. This work was also supported by TUBITAK (Project Number; 100082) and Ministry of Development, Republic of TURKEY (Project Number; 100139).
\end{acknowledgements}




\end{document}